# Type of the Paper (Editorial)

# A Health Focused Text Classification Tool (HFTCT)

**Baadr Suleman M Alwheepy[1], Leandros Maglaras[1] and Nick Ayres[1]**

[1] Cyber Technology Institute ; De Montfort University ; leandros.maglaras@dmu.ac.uk

**Abstract:** Due to the high number of users on social media and the massive amounts of queries requested every second to share a new video, picture, or message, social platforms struggle to manage this humungous amount of data that is endlessly coming in. HFTCT relies on wordlists to classify opinions. It can carry out its tasks reasonably well; however, sometimes, the wordlists themselves fail to be reliable as they are a limited source of positive and negative words.

## 1. Introduction

Twitter is a social platform where users share their thoughts and opinions on various topics (Forsey, 2019); It is a platform with nearly 396.5 million users from all around the world (Dean, 2022). Throughout this project, a Health-Focused Text Classification tool (HFTCT) was under development; it is a tool that performs sentiment analysis and opinion classification on Twitter. It classifies opinions on any subject of the user's choice and concludes Twitter users' opinions in a short report showing the positive to negative ratio of the tweets surrounding that particular topic.

The primary data source for HFTCT is the tweets that are endlessly being shared on Twitter. It is a very liable source for the tool, mainly because these opinions can be shared from anyone worldwide, making this source very multicultural and broad as tweets could cover various ongoing events and interesting topics from all around the globe.

The motivation that led to developing HFTCT was encouraged by a problem that exists in nearly every social platform. Due to the high number of users on social media and the massive amounts of queries requested every second to share a new video, picture, or message, social platforms struggle to manage this humungous amount of data that is endlessly coming in.

This type of data is called Big Data; it is large sets of structured and unstructured information (SAS, n.d.; Russom, 2011). Big data is a problem for social networks because it builds up quickly and takes so much storage space. The continuous flow of data pressures many social platforms to store the data shared in an unstructured manner, creating big chunks of very complex data that is very tough to analyse.

The process of extracting useful information out of these vast datasets is called big data analysis; it is a very beneficial process, primarily because it turns unanalysed data chunks that many may consider worthless into something meaningful (Abdul Ghani et al., 2019; Chai, 2021). This process is carried out by big data analytics tools such as HFTCT, and if developed to a great extent, they can highlight insights and relationships within these enormous datasets.

HFTCT aims to assist medical organisations, researchers, and medical professionals by providing a short and accurate report summarising what the public thinks of any health-related topic. This summary can be very beneficial for professionals; it would give them a view from a different standpoint, help them acknowledge what causes the negativity, and potentially assist them in clarifying any misconceived thoughts shared on Twitter.

The tool should save its users so much time as it would provide that summary within seconds. It could potentially be put into many other uses; for example, it can be used for hospital reviews; a user can enter a hospital name and would be able to get a summary of what the public thinks of that particular hospital.

**2. Related Work**

The research within technical aspects has shown that many tools that perform in-text classification and sentiment analysis were previously developed; however, most of these tools focused on assisting professionals in political and business-related fields. The literature review highlights the lack of implementation of such tools within medical fields. It emphasises how vital and beneficial these tools can be for medical organizations, especially through the ongoing medical crisis (covid-19) and the hesitancy to take covid vaccines (Alwheepy, 2022).

On the other hand, the research on the physiological aspects gave a better understanding of how human beings express their emotions through verbal and non-verbal expressions. Moreover, throughout the research in physiological fields, many emotion classification frameworks were developed by psychologists; these frameworks help identify a person's emotional state and to what extent they may develop (Alwheepy, 2022; Izard, 2007).

Overall, the literature review suggests that sentiment analysis tools are highly beneficial and save so much time for professionals in various fields. Finally, it also suggests that the use of such tools within medical fields can help medical organisations and researchers with many matters, one of which is reducing the hesitancy to take the covid vaccine. Professionals would be able to define the reasons behind the refusal and find better approaches to convincing people who are hesitating to take the vaccine. Resolving this hesitancy could lead to many social benefits as the safer a community is, the more socially active it becomes and the less likely it is to be put under lockdown. This could also contribute to economic and educational benefits as the medical crisis would not interrupt businesses or educational institutions.

2.1 Survey

Along with the literature review, more research was carried out recently; a survey on social media use was conducted to acknowledge how social platforms are used differently across different age groups. The study involved 327 participants who answered the survey questions; the questions were focused on identifying what age group is more likely to share their thoughts and opinions on social media and whether these thoughts were mostly positive or negative.

As for the methods, the survey was conducted on a wide range of age groups; under-18's, 18-25, 26-35, 36-45, and +45 years old. It was posted on multiple social platforms as the intention was to assess social media users; it was also conducted on people who were met in face-to-face gatherings. Chart 1 below shows the examined participants' age and as can be seen from the chart, most of the assessed induvial were aged between 18 and 25.

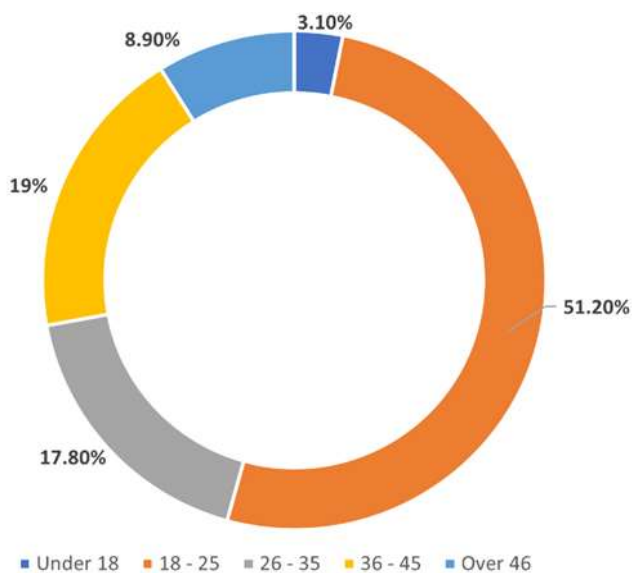

Chart 1 – showing the participants' age groups

As mentioned above, one of the survey's goals was to determine the difference in social media use across different age groups; therefore, one of the starter questions asked the participants about the number of daily hours they spend on social media; the results have shown that 42.9% spend over 6 hours daily on social media, and 39.9% spend over 3 hours; which leaves 17.2% who either spend less than 3 hours a day or do not use social media on a daily basis. Graph 1 below shows the discussed data with higher detail levels; it shows each age group's daily number of hours on social media platforms. The statistics in Graph 1 prove how important social media is in our lives; the data shows that 50% of the people who are 18 to 25 years old spend over six hours daily on social media, which is the equivalent of nearly 91 days each year.

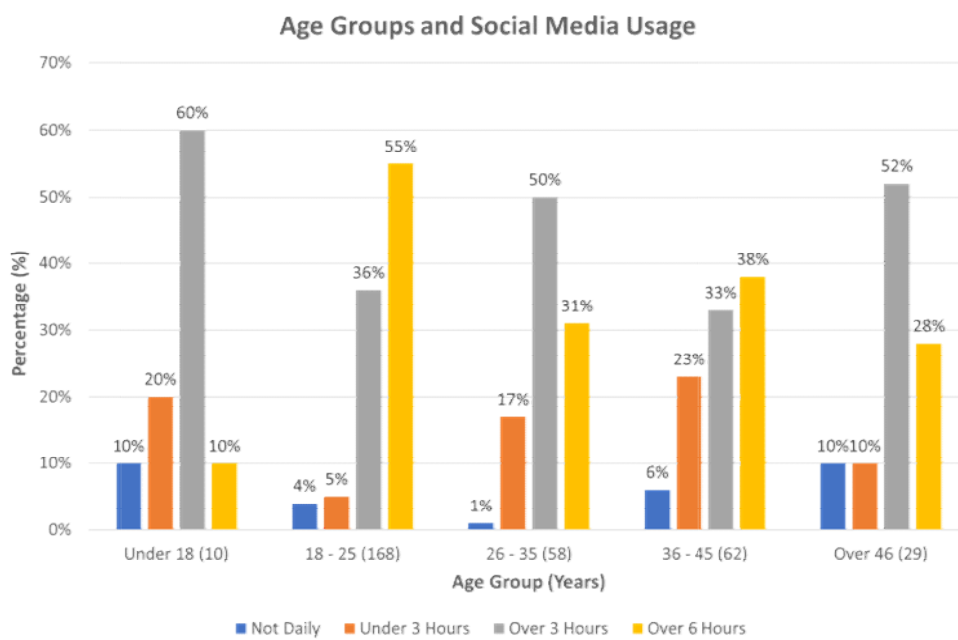

Graph 1 – Showing the number of hours spent on social media by each age group

Note: the number between the brackets is the number of surveyed induvial from that particular age group.

Moreover, one of the questions that were asked in the survey was how often each person would share their thoughts/opinions on social platforms. The results have shown that 22.7% of all 327 participants regularly share their opinions on social platforms; it has also shown that 51.5% said that they do it rarely, and this leaves the last 25.8% who claim that they never do. In order to take these findings a step further, all participants were asked whether these shared thoughts were mostly positive or negative; these results are displayed in Graph 2, which shows the difference in the posts shared between each age group.

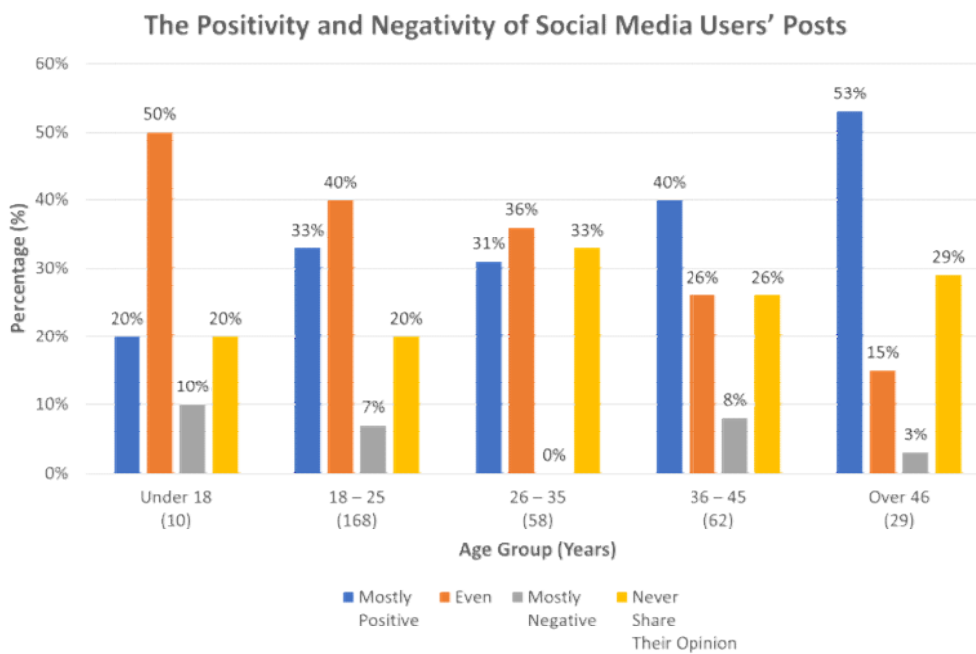

Graph 2 – Showing the difference in the posts shared between each age group

Note: the number between the brackets is the number of surveyed induvial from that particular age group.

The statistics in Graph 2 above show that the people who mostly share negative posts on social media are very few compared to the other categories; it also shows that all age groups have a high percentage of people who choose not to share their thoughts/opinions on social platforms which indicates that there is no relationship between age and opinion sharing. Finally, Graph 2 also shows a clear relationship between age and positivity; the graph indicates that the older a person is, the more likely they are to share a positive post.

Lastly, the survey also asked participants whether they feel like their opinions can make environmental, social, or economic impacts. Chart 2 below illustrates the results; it shows that 53.7% of all participants believed that their opinions can make an impact but not always; 33.7% believed that their opinions definitely make an impact, and the last 12.6% believed that they make no impact at all.

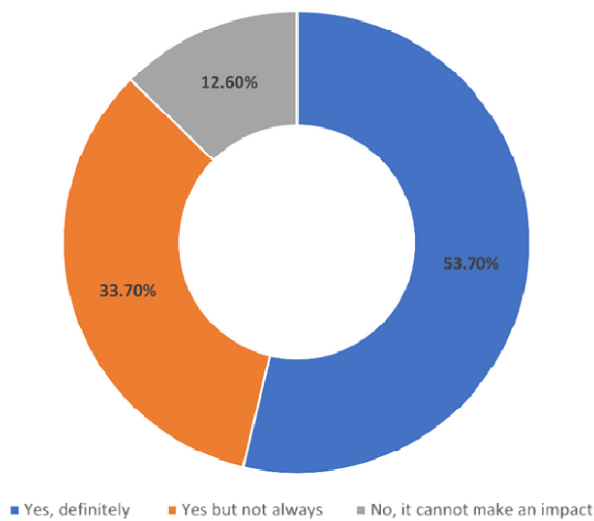

Chart 2 – showing the participants about the impacts of their opinions

The statistics on Chart 2 can change with the help of in-text classification tools like HFTCT because professionals in any field would be able to run a classification against any topic on social media platforms and be able to acknowledge why people are positive or negative about certain matters. In other words, there would be a higher chance for peoples' voices to be heard with the use of such tools, as every shared opinion can easily be accessed and reviewed by a professional. Finally, this could also encourage people to share their opinions more often and could later on result in resolving matters at a faster pace.

3. Tool Components:

HFTCT's components are located on both the front-end and the back-end of the application. It consists of 5 major components, a user interface, Twitter API queries, filter, classification system, and a wordlist. These elements are all essential for the tool to function; they are all developed to perform or participate in various tasks that help provide a better classification result for the user. Figure 1 below illustrates the application's workflow and the order in which every component performs.

Figure 1 – HFTCT's workflow

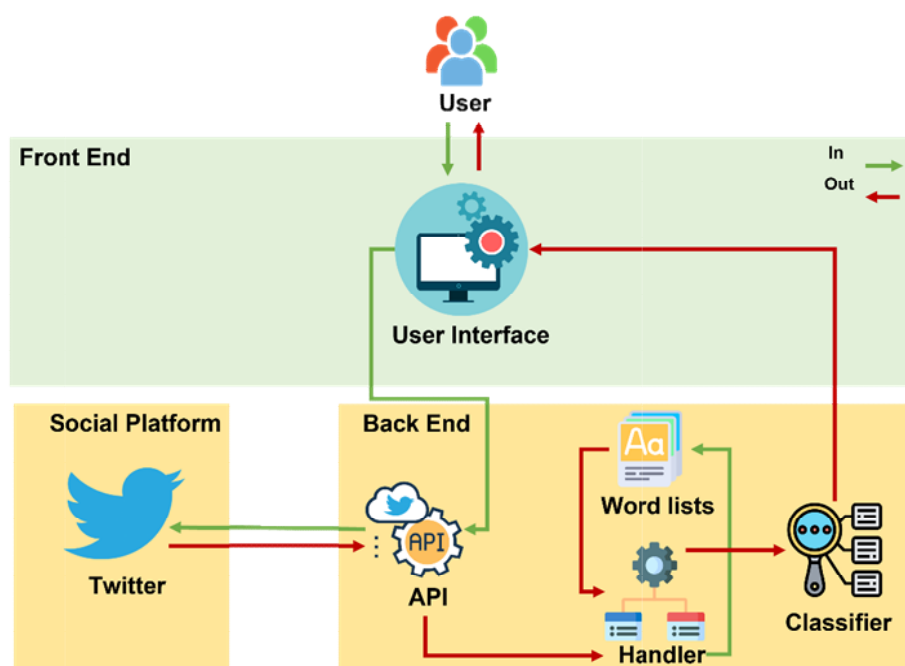

3.1 Graphical User Interface

The graphical user interface is the only front-end component; it is the element that interacts with the user through a window and links their interaction with the back-end components. The primary goal of designing and developing a GUI is to make an application more presentable. A GUI also enhances the tool's usability and makes it more accessible; this is highly essential, primarily because HFTCT aims to assist those working within the medical field, and some of them may find running the tool through a terminal challenging (Haq, 2022).

Image 1 below shows the GUI's main window and the used layout while Figure 2 illustrates how user request is passed from

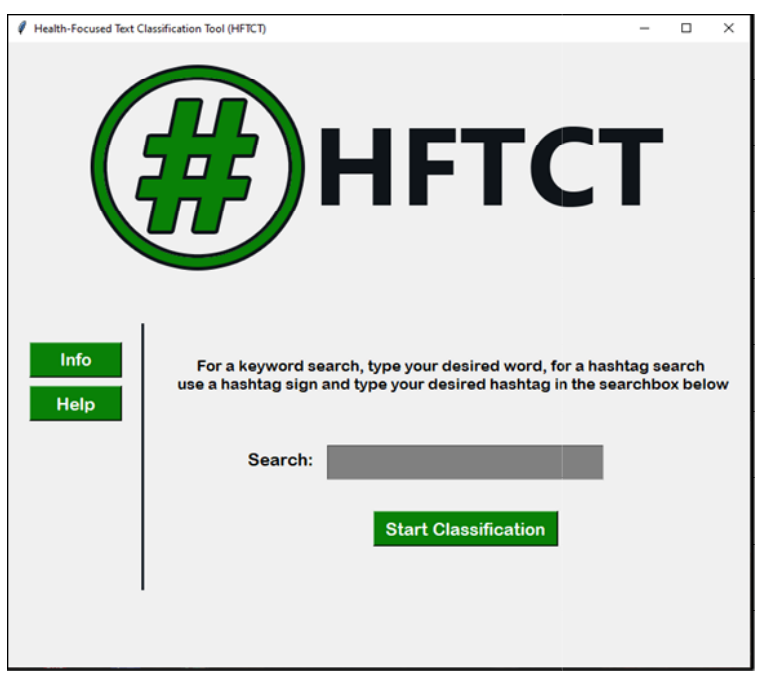

Image 1 – HFTCT's main window the front-end to the back-end.

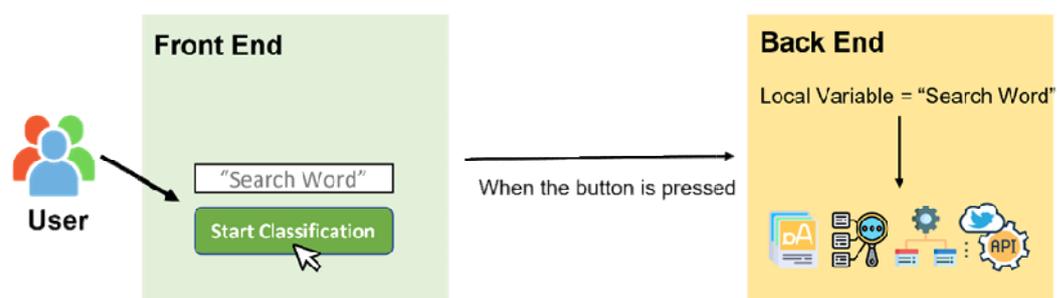

Figure 2 – Front and back end communication

### 3.2 Twitter API and User Querries

Twitter API is a publicly accessed component used in the back-end of HFTCT. Gaining access to Twitter API was one of the early-stage challenges; it is a component of very high importance in HFTCT because it connects the application to the social network. In order to use the API, a Twitter developer account is needed; the account gives its users the privileges to access the data on the platform and use it on their projects.

After establishing a connection to Twitter's social network, the component can be used to search for tweets based on the user's desired keyword or hashtag. When users write their desired search in the tool, their input gets stored in a string variable named "SearchWord". The reason behind storing the input in a string variable is to give the user the space and freedom to write whatever they desire from numbers, a single word, or even multiple words, as they would all be accepted within a string variable.

When the component sends a search query, the response of that query is and then passed into a for loop that arranges the tweets by putting every tweet as part of a list of strings where every item in the list is an entire tweet. Figure 3 illustrates how tweets are stored.

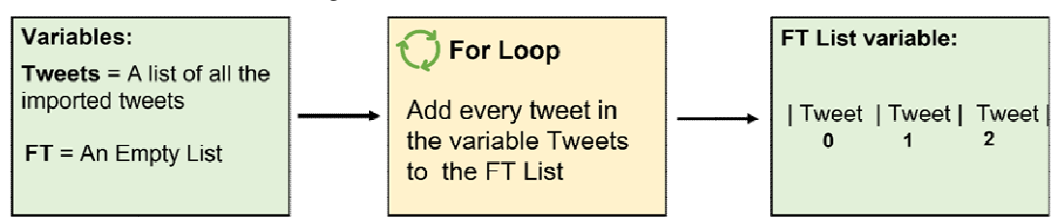

Figure 3 – How tweets are stored

### 3.3 Tweet Handler and Filter

Unlike formal documentation, Twitter tweets and all other social networks' posts tend to be informal; they usually contain many spelling errors, internet slang, and sentences that contain grammar issues (Chatterjee et al., 2018). These informalities are a challenge to all in-text classification systems as they make automatic analysis harder; however, this component aims to ease up the structure of the imported tweets by organising them, correcting and filtering out some mistakes.

HFTCT's classification system relies on wordlists; it classifies the imported input by comparing every word against a positive and a negative word list, which explains the reasoning behind the first objective. Figure 4 illustrates the method works; it splits depending on whitespace, meaning whenever it finds whitespace within a string, it will split it and store it as an item in the list.

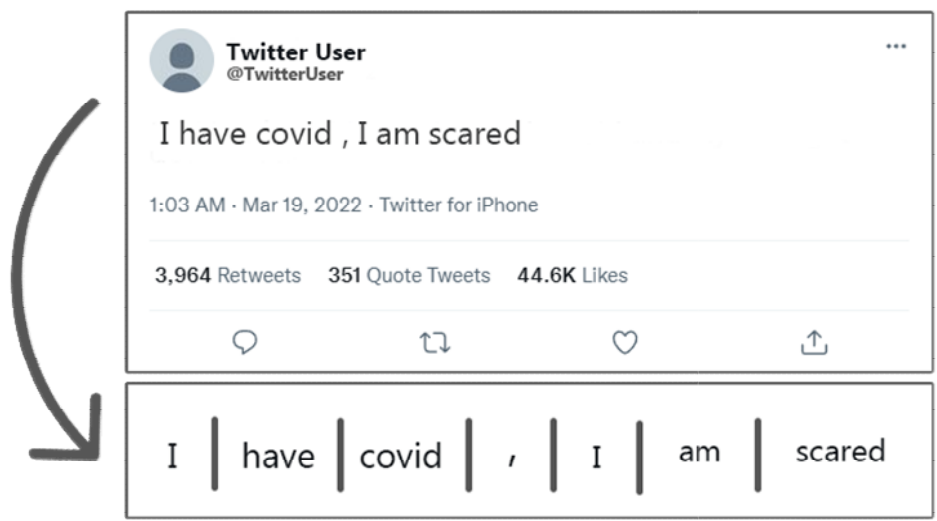

Figure 4 – How the handler splits tweets store them in a list variable

Finally, after splitting a tweet word by word, every word gets passed into two if statements. The first if statement checks whether the word exists in the negative/positive word list, and if it was found, then the program adds a value of one to the count. The second if statement checks whether the word before the negative/positive word is a term that changes the meaning completely; for example, in a case where a Twitter user types "I am not sad", this smart system would be able to detect that the user is not sad and would not misclassify that tweet. Figure 5 below illustrates the functionality of this handler.

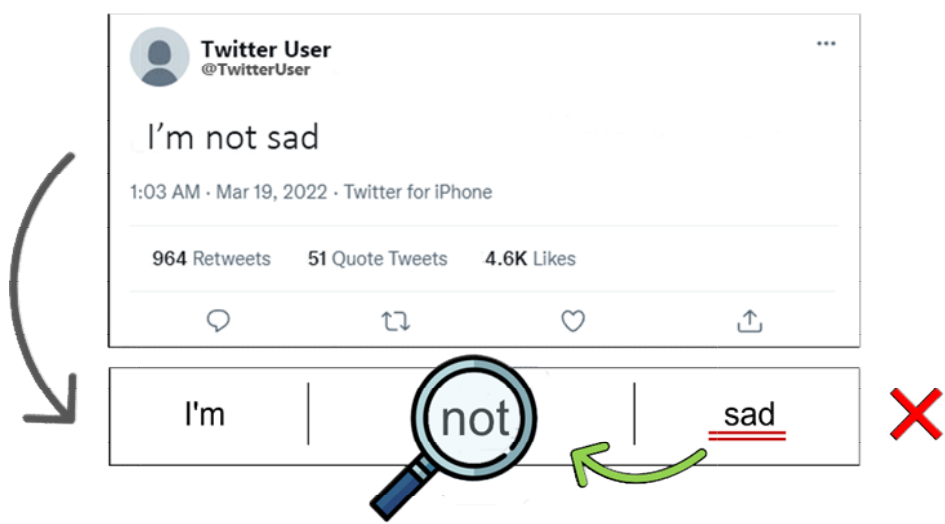

Figure 5 – showing how the reverse term functionality functions

3.4 Classification System

The classifier in HFTCT is a binary classification system; it relies on the positive and negative count values stored in the handler component to perform a calculation of the positivity and negativity percentages on the user's chosen topic. The component follows Formulas 1 and 2; it simply calculates the positive/negative percentage by dividing a hundred by the total number of words found within the wordlists and multiplying the answer with the positive/negative count. The result of the calculation is then stored in a variable to be later displayed on the user interface for the user.

$$Positivity\ Percentage = \frac{100}{Positive\ Words\ Found\ +\ Negative\ Words\ Found}\ X\ Positive\ Count$$

Formula 1 – The positivity percentage formula

$$Negativity\ Percentage = \frac{100}{Positive\ Words\ Found\ +\ Negative\ Words\ Found} \times Negative\ Count$$

Formula 2 – The negativity percentage formula

The program allows a user to store his/her results in an CSV file that contains more information about the tweet and the user themselves, for example, the time and date of when the tweet was posted, the username, the tweet itself, and the negative/positive words found within the tweet. This functionality was added to give the user the ability to check and understand the reasoning behind the positivity/negativity around their topic, as well as help them identify where certain opinions are located and during what time events they were posted.

3.5 Word lists

This component consists of three lists, a positive word list, a negative word list, and a word list for the reversed terms functionality used in the handler. In total, these word lists contain approximately 7000 words, most of which are taken from a sentiment analysis project that was developed in 2004 (Hu and Liu, 2004; Cheng, Hu, and Liu, 2005). These word lists are a little old but reliable; they have been slightly amended throughout the development of this component and more words were added to the lists. The primary goal of this component is to be always available for the program when it runs.

The development of all of the components of HFTCT was done in a particular order; this is shown in Figure 6 below. Following a development order was necessary as some components rely on each other; for example, the tweet handler relies on the tweets imported through the Twitter API query; therefore, the API component should ideally be developed first.

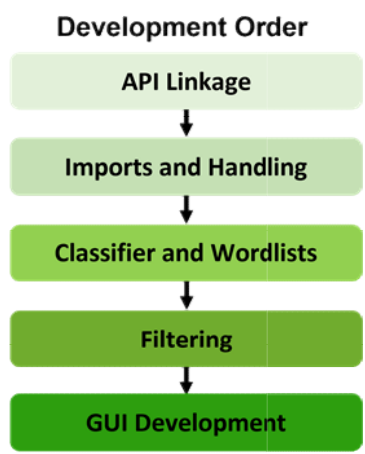

Figure 6 – showing the development order

4. Evaluation

In order to evaluate the performance of the tool we conducted several tests that included the classification of positive and negative works, the display of classification results and the tweets.

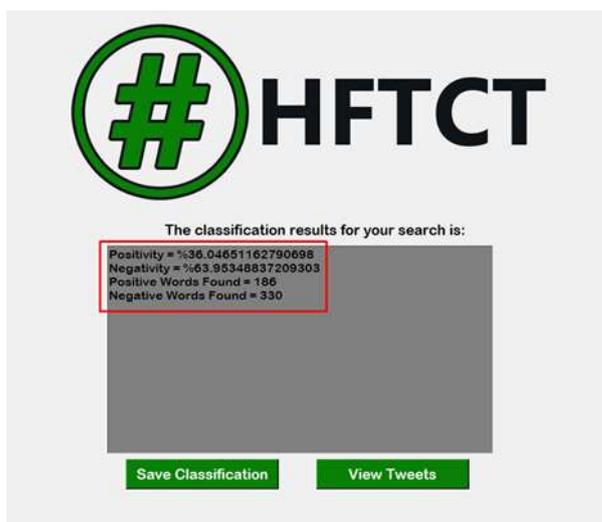

Screenshot 1 – showing the calculation results performed by the tool

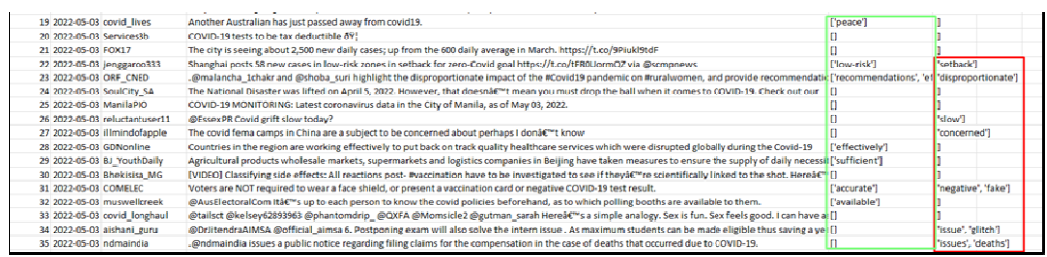

Screenshot 2 – Showing a saved file of a classification results that displayed the detected positive/negative words

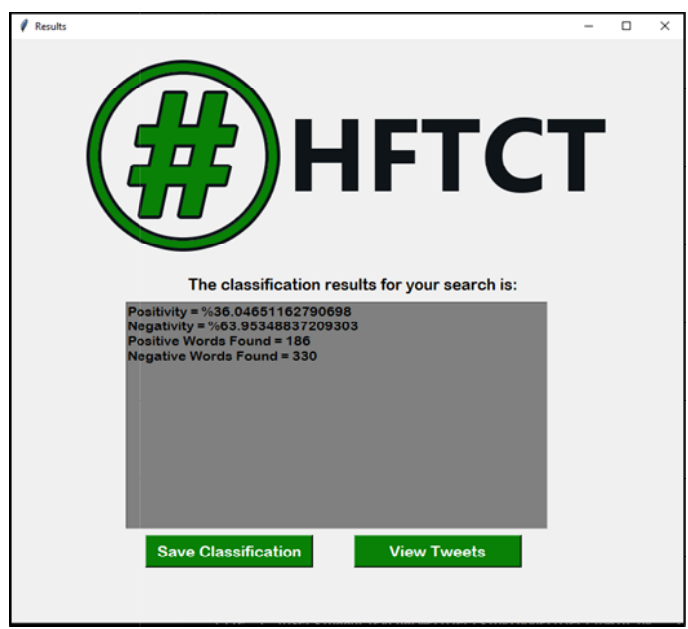

Screenshot 3 – showing the program's ability to display the classification result

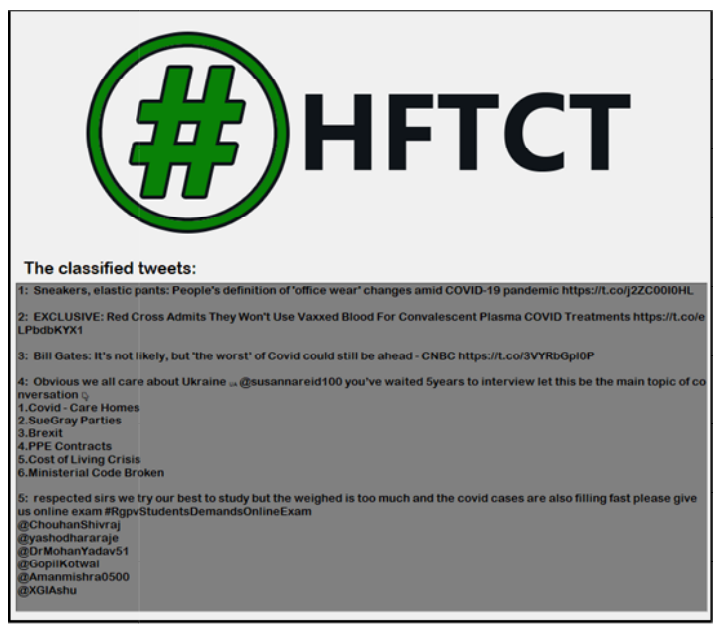

Screenshot 4 – showing the program's ability to display the imported tweets

5. Discussion - conclusions

As discussed previously in this paper, HFTCT relies on wordlists to classify opinions; it can carry out its tasks reasonably well; however, sometimes, the wordlists themselves fail to be reliable as they are a limited source of positive and negative words. In other words, basing the classification on wordlists is practical but may sometimes affect the accuracy due to the limitations of the source, which in this case is the wordlists. A solution that could potentially tackle this issue entirely is to have a machine learning functionality that constantly updates the wordlist. This solution can be applied by adding a functionality that checks for any unidentified words within a text, then searches that word across multiple online libraries such as Urban Dictionary, which is a library that defines any newly invented slang words. What makes this solution machine learning is that when an unidentified word is found to be negative or positive, this functionality would add it to the wordlist.

Several ideas for further enhancements are planned to be implemented in the future. One of these ideas is to develop the classifier even further to an extent where it recognises patterns within a text and identifies human emotions. Applying this will undoubtedly require more research on the psychological aspects of human emotions; however, it would certainly make the tool more beneficial.

Another idea is to add a feature that allows the user to search within a certain period of time. The idea of this feature is to allow users to compare two or more classifications about the same topic but from different time periods; for example, with this feature, researchers would be able to classify tweets that were posted at the early stages of the covid-19 pandemic and compare the results with a classification of recent tweets. This feature would be very beneficial for the user as it would allow them to see the progression of human emotions during any event that happened previously.

Finally, another feature that is planned to be implemented in the future is the idea of importing tweets from a specified geographic location. Similarly, with this feature implemented, researchers would be able to compare classification results from various geographic locations. This feature can be very beneficial as the user would be able to visualize the effect of environmental factors such as temperature, population density, and pollution on human emotions.

In conclusion, although HFTCT can achieve its grand goal and assist the targeted audience with its classifications, it is essential to acknowledge that the classifier's accuracy is adequate but still requires some improvements due to the obstacles it faces.

**Funding:** This research received no external funding

**Conflicts of Interest:** The authors declare no conflict of interest